\documentclass[aps,prb,twocolumn]{revtex4} 

\usepackage{graphicx}
\usepackage{dcolumn}
\usepackage{bm}
\usepackage{amsmath} 

\newcommand{\comment}[1]{}

\begin{document}

\title{Quantum Monte Carlo in Classical Phase Space
with the Wigner-Kirkwood Commutation Function. II.
Diagonal Approximation in Position Space.}


\author{Phil Attard}
\affiliation{ {\tt phil.attard1@gmail.com}  13 Jan.\ 2026, \today }


\begin{abstract}
A third order expansion for Wigner-Kirkwood commutation function,
a complex function in classical phase space
that accounts for the Heisenberg uncertainty relation,
is approximated and integrated over momentum
to give a real function in position configuration space.
Metropolis Monte Carlo computer simulation results are given
for liquid Lennard-Jones $^4$He below 10\,K.
\end{abstract}


\maketitle

%
\section{Introduction}
\setcounter{equation}{0} \setcounter{subsubsection}{0}
\renewcommand{\theequation}{\arabic{section}.\arabic{equation}}
%

This paper follows on from Attard (2025h),
which itself is based on earlier work (Attard 2016b, 2018b, 2021).
The immediately preceding work presented computer simulation results
in classical phase space
for saturated liquid $^4$He near the $\lambda$-transition,
with certain quantum effects (ie.\ the Heisenberg uncertainty relation)
included
via a third order expansion of the Wigner-Kirkwood commutation function,
which is both complex and a function of the momentum configuration.
The purpose of the present paper
is to explore an approximation that reduces
the latter to a real function of position configuration space.
There are certain practical and conceptual advantages to this,
and there is apparently little loss of accuracy.

%
\section{Formalism and Algorithm}
\setcounter{equation}{0} \setcounter{subsubsection}{0}
\renewcommand{\theequation}{\arabic{section}.\arabic{equation}}
%

\subsection{Phase Space Weight}

For a subsystem of  $N$ identical bosons,
a point in classical phase space is
${\bf \Gamma} = \{{\bf q},{\bf p}\}$.
Here the position configuration is
${\bf q} = \{ {\bf q}_1,{\bf q}_2,\ldots, {\bf q}_N\}$
and the momentum configuration is
${\bf p} = \{ {\bf p}_1,{\bf p}_2,\ldots, {\bf p}_N\}$,
where the position of boson $j$ is
${\bf q}_j = \{ {q}_{jx},{q}_{jy},{q}_{jz}\}$,
and its momentum is
${\bf p}_j = \{ {p}_{jx},{p}_{jy},{p}_{jz}\}$.
Here we take the momentum to belong to the continuum;
see Attard (2025a, 2025b) for the treatment of quantized momentum.

For a canonical equilibrium system of temperature $T$ and volume $V$,
the phase space probability density is
(Attard 2018b, 2021)
\begin{equation}
\wp({\bf \Gamma}) =
\frac{
e^{-\beta {\cal H}({\bf \Gamma})}
e^{W({\bf \Gamma})} \eta({\bf \Gamma})
 }{ N! h^{3N}Z },
\end{equation}
where $h$ 
is Planck's constant,
and $\beta = 1/k_{\rm B}T$,
$k_{\rm B}$ 
being Boltzmann's constant.
The partition function, $Z(N,V,T)$,
normalizes the probability and its logarithm gives the total entropy.
Also the classical Hamiltonian is
${\cal H}({\bf \Gamma}) = {\cal K}({\bf p}) + U({\bf q})$,
where ${\cal K}({\bf p}) = p^2/2m = \sum_{j=1}^N p_j^2/2m$
is the kinetic energy,
$m$ being the mass of a boson.
Below we shall take the potential energy
to consist of central pair potentials
$U({\bf q}) = \sum_{j<k}^N u(q_{jk})$
and write $u_{jk}  \equiv u(q_{jk})$.


The main focus of this work is on the Wigner-Kirkwood
commutation function $W({\bf \Gamma})$
(Attard 2018b, 2021, Kirkwood 1933, Wigner 1932),
which is defined by
\begin{equation} \label{Eq:Wdefn}
e^{-\beta {\cal H}({\bf \Gamma})}
e^{W({\bf \Gamma})}
=
e^{{\bf p}\cdot{\bf q}/{\rm i} \hbar}
e^{-\beta \hat{\cal H}({\bf q})}
e^{-{\bf p}\cdot{\bf q}/{\rm i} \hbar} .
\end{equation}
This corresponds to $\omega_p =e^{W_p}$ (Attard 2018b Eq.~(2.6))
(present notation).
For the quantum weight one can use either $\omega_p \eta_q$
or else $\omega_q \eta_p$,
with $\omega_q^* = \omega_p$ and $\eta_q^* = \eta_p$.

On the right hand side of the definition of the commutation function
the Hamiltonian operator appears,
$ \hat{\cal H}({\bf q}) =  \hat{\cal K}({\bf q}) +  U({\bf q})$.
The Fourier factors are unnormalized, unsymmetrized momentum eigenfunctions.
The non-zero commutator $[\hat{\cal K}, U] \ne 0$
is what makes $W \ne 0$,
and so it reflects the Heisenberg uncertainty relation.


In the computational results for $^4$He presented below
the symmetrization function is neglected.
This is appropriate on the high temperature side of the $\lambda$-transition.
Our main aim is to delineate the r\^ole of the commutation function,
and we shall take up the combined effects of the two in future work.


\subsection{Fluctuation Expansion for the Commutation Function}

Kirkwood (1933) took the inverse temperature derivative
of the defining equation for the commutation function, Eq.~(\ref{Eq:Wdefn}),
to obtain an infinite series in powers of $\beta$,
explicitly giving the first non-zero term, which is quadratic in $\beta$.
Higher order coefficients using this method have been obtained
(Attard 2021 \S\S8.2--8.4).

A slightly different approach based on a fluctuation expansion
has $n$th order coefficient (Attard 2021 \S8.5)
\begin{equation}
\Delta_{\cal H}^{(n)}({\bf q},{\bf p})
\equiv
\frac{ \langle {\bf q} | \big[ \hat{\cal H}
- {\cal H}({\bf q},{\bf p}) \big]^n | {\bf p} \rangle
}{ \langle {\bf q} |  {\bf p} \rangle }  .
\end{equation}
One has $\Delta^{(0)}_{\cal H}({\bf q},{\bf p}) = 1$
and $\Delta^{(1)}_{\cal H}({\bf q},{\bf p}) = 0$.
These give a series for the exponent of the form
\begin{equation}
W(\Gamma) =
\frac{\beta^2}{2!} \tilde \Delta_{\cal H}^{(2)}(\Gamma)
-\frac{\beta^3}{3!} \tilde \Delta_{\cal H}^{(3)}(\Gamma)
+ \frac{\beta^4}{4!} \tilde \Delta_{\cal H}^{(4)}(\Gamma)
-\ldots
\end{equation}
The second order fluctuation,
$\tilde \Delta^{(2)}_{\cal H} = \Delta^{(2)}_{\cal H}$, is
\begin{eqnarray}
\Delta^{(2)}_{\cal H}({\bf q},{\bf p})
 & = &
\frac{-\hbar^2}{2m} \nabla^2 U({\bf q})
 - \frac{\mathrm{i}\hbar}{m} {\bf p} \cdot \nabla  U({\bf q}) .
\end{eqnarray}
The third order fluctuation,
$\tilde \Delta^{(3)}_{\cal H} = \Delta^{(3)}_{\cal H}$, is
\begin{eqnarray}
\Delta^{(3)}_{\cal H}({\bf q},{\bf p})
& = &
\frac{- \hbar^2}{m}  \nabla U\! \cdot\!  \nabla U
+ \frac{\hbar^4}{4m^2} \nabla^2 \nabla^2 U
 \\ && \mbox{ }\nonumber
+ \frac{\mathrm{i}\hbar^3}{m^2} {\bf p} \!\cdot\! \nabla \nabla^2 U
- \frac{\hbar^2}{m^2} {\bf p} {\bf p}  : \nabla \nabla  U .
\end{eqnarray}
Expressions for the gradients of a central pair potential
have been catalogued
(Attard 2021 \S\S 9.5.2 and 9.5.3).


\subsection{Momentum Integral in Diagonal Approximation}

\subsubsection{Pair Commutation Function}

The real part of the second-order commutation function,
which is independent of momentum, is
\begin{eqnarray}
\tilde W^{(2)}_{\rm r}({\bf q})
& = &
\frac{\beta^2}{2}
\frac{-\hbar^2}{2m}  \nabla^2 U({\bf q})
\nonumber \\ & = &
\frac{-\beta^2\hbar^2}{4m}
\sum_{j<k} [ \nabla^2_j  u_{jk} + \nabla^2_k  u_{jk} ]
\nonumber \\ & = &
\frac{-\beta^2\hbar^2}{4m} \sum_{j,k} \!^{(k \ne j)}
\left[ u''_{jk} + \frac{2u'_{jk}}{q_{jk}} \right] .
\end{eqnarray}
Here and below $u'_{jk} = \partial u(q_{jk})/\partial q_{jk}$,
$u''_{jk} = \partial^2 u(q_{jk})/\partial q_{jk}^2$,
and similarly for higher order derivatives.

The imaginary momentum part of the second order fluctuation
contributes
\begin{equation}
e^{{\rm i} W^{(2)}_{\rm i}}
=
e^{{\rm i}\sum_{j=1}^N W^{(2)}_{{\rm i},j}/2},
\end{equation}
where the part involving boson $j$ is
\begin{eqnarray}
W^{(2)}_{{\rm i},j}
& = &
\frac{-\beta^2\hbar}{2m}   [ {\bf p} \cdot \nabla  U({\bf q}) ]_j
\nonumber \\ & = &
\frac{-\beta^2\hbar}{2m}  \sum_{k=1}^N \!^{(k \ne j)}
[ {\bf p}_j \cdot \nabla_j u_{jk} + {\bf p}_k \cdot \nabla_k u_{jk} ]
\nonumber \\ & = &
\frac{-\beta^2\hbar}{2m}  \sum_{k=1}^N \!^{(k \ne j)}
[ \frac{{\bf p}_j \cdot {\bf q}_{jk} }{q_{jk}}  u'_{jk}
+  \frac{{\bf p}_k \cdot {\bf q}_{kj} }{q_{kj}}  u'_{kj} ]
\nonumber \\ & = &
\frac{-\beta^2\hbar}{2m}  \sum_{k=1}^N \!^{(k \ne j)}
\frac{{\bf p}_{jk }\cdot {\bf q}_{jk} }{q_{jk}}  u'_{jk}.
\end{eqnarray}
This has to be divided by two in the total
because it counts each pair interaction twice when summed over $j$.
Writing the total dependence on $j$ is useful
for the Metropolis Monte Carlo step.

We can write the total of this contribution as
\begin{eqnarray}
W^{(2)}_{\rm i}
& = &
\prod_{j<k}
e^{ (-{\rm i}\beta^2\hbar u'_{jk} /2mq_{jk})
{\bf p}_{jk }\cdot {\bf q}_{jk} }
\nonumber \\ & = &
\prod_{j<k}
e^{ (-{\rm i}\beta^2\hbar u'_{jk} /2mq_{jk})
[ {\bf p}_{j }\cdot {\bf q}_{jk}  + {\bf p}_{k }\cdot {\bf q}_{kj} ] }
\nonumber \\ & = &
\prod_{j,k} \!^{(k \ne j)}
e^{ (-{\rm i}\beta^2\hbar u'_{jk} /2mq_{jk}) {\bf p}_{j }\cdot {\bf q}_{jk} }
\nonumber \\ & = &
\prod_{j}
e^{ -{\rm i}{\bf a}_{j} \cdot {\bf p}_{j}/ \hbar } ,
\end{eqnarray}
where the corresponding `position' is
\begin{equation}
{\bf a}_{j}({\bf q}) \equiv
\frac{ \beta^2\hbar^2  }{ 2 m}
\sum_{k=1}^N \!^{(k \ne j)}
\frac{ u'_{jk} }{ q_{jk}} {\bf q}_{jk} .
\end{equation}
The magnitude of each pair contribution to this
is large in the core region,
and it goes to zero at large separations.
With it the momentum integral is
\begin{eqnarray}
I_{W2} & = &
\frac{1}{h^{3N}} \int {\rm d}{\bf p}\; e^{-\beta p^2/2m}
\prod_{j=1}^N e^{ - {\bf a}_{j} \cdot {\bf p}_{j}/{\rm i}\hbar }
\nonumber \\ & = &
\frac{1}{h^{3N}}  \prod_{j=1}^N \int {\rm d}{\bf p}_j \;
e^{-\beta p_j^2/2m}
e^{ -{\bf a}_{j} \cdot {\bf p}_{j}/{\rm i} \hbar }
\nonumber \\ & = &
\Lambda^{-3N} \prod_{j=1}^N e^{ -\pi a_j({\bf q})^2 /\Lambda^2 } .
\end{eqnarray}
The thermal wavelength is $\Lambda = \sqrt{2\pi\hbar^2\beta/m}$.
Since $a_j$ is large at small separations,
this keeps particle $j$ more separated from its neighbors
than in the classical case,
which is to say in the absence of the Wigner-Kirkwood commutation function.

\subsubsection{Triplet Commutation Function}

For the third order fluctuation,
there are two terms that are independent of momentum.
The first is
\begin{eqnarray}
\tilde W^{(3)}_{{\rm r},a}
& = &
\frac{-\beta^3}{3!}  \frac{-\hbar^2}{m}
\nabla U\! \cdot\!  \nabla U
\nonumber \\ & = &
\frac{\beta^3\hbar^2}{3!m} \sum_{j=1}^N
\sum_{k,k'}  \nabla_j u_{jk} \cdot \nabla_j u_{jk'}
\nonumber \\ & = &
\frac{\beta^3\hbar^2}{3!m} \sum_{j=1}^N {\bf g}_j \cdot {\bf g}_j,
\end{eqnarray}
where
\begin{equation}
{\bf g}_j
\equiv \sum_{k=1}^N \!^{(k \ne j)} \nabla_j u_{jk}
= \sum_{k=1}^N \!^{(k \ne j)} \frac{u_{jk}'}{q_{jk}} {\bf q}_{jk}.
\end{equation}
Note that ${\bf g}_j$ depends upon ${\bf q}_j$ and also ${\bf q}_k$
for $k$ within the potential cut-off neighborhood of $j$.
This must be taken into account in the Metropolis Monte Carlo step.

The second real momentum-independent part is
\begin{eqnarray}
\tilde W^{(3)}_{{\rm r},b}
& = &
\frac{-\beta^3}{3!}  \frac{\hbar^4}{4m^2} \nabla^2 \nabla^2 U
\nonumber \\ & = &
\frac{-\beta^3\hbar^4}{12m^2}
\sum_{j,k}\!^{(k \ne j)}
\left\{ u^{\rm iv}_{jk}  + \frac{4u'''_{jk}}{q_{jk}} \right\} .
\end{eqnarray}

The momentum-dependent terms in the third order fluctuation
are an imaginary term linear in momentum,
$ ({\mathrm{i}\hbar^3}/{m^2}) {\bf p} \!\cdot\! \nabla \nabla^2 U$
and a real term that is quadratic in momentum
$(- {\hbar^2}/{m^2}) {\bf p} {\bf p}  : \nabla \nabla  U$.
These are multiplied by $-\beta^3/3!$.

In time-honored tradition,
for the quadratic momentum term
we neglect the `difficult' off-diagonal terms, ${\bf p}_{j} {\bf p}_{k} $.
The justification is that these can be positive or negative
and therefore they largely cancel.
In fact the first neglected term that is non-zero
on average is ${\cal O}(p^4)$,
which is small at low temperatures,
more so in the quantum than the classical case.
In view of this we keep only the leading terms even in momentum,
which gives the diagonal approximation, namely
\begin{eqnarray} \label{Eq:pdiag}
W^{(3)}_{{\rm r},c}
& = &
\frac{-\beta^3}{3!} \frac{-\hbar^2}{m^2} {\bf p}{\bf p} : \nabla\nabla U
\nonumber \\ & \approx &
\frac{\beta^3\hbar^2}{6m^2}
\sum_{j,\alpha} p_{j\alpha}^2 \partial_{j\alpha}^2 U
\nonumber \\ & = &
\frac{\beta^3\hbar^2}{6m^2}
\sum_{j,\alpha} \sum_k\!^{(j\ne k)}
p_{j\alpha}^2 \partial_{j\alpha}^2 u_{jk}
\nonumber \\ & = &
\frac{\beta^3\hbar^2}{6m^2}
\sum_{j,\alpha} \sum_k\!^{(j\ne k)} p_{j\alpha}^2
\nonumber \\ && \mbox{ } \times
\left[  \frac{u''_{jk} q_{jk,\alpha}^2}{q_{jk}^2}
-
 \frac{u'_{jk} q_{jk,\alpha}^2}{q_{jk}^3}
 +
 \frac{u'_{jk}}{ q_{jk}}  \right]
\nonumber \\ & \equiv &
\sum_{j=1}^N \sum_{\alpha=x,y,z}
B_{j\alpha}({\bf q}) p_{j\alpha}^2 ,
\end{eqnarray}
where
\begin{eqnarray}
B_{j\alpha}({\bf q})
& \equiv &
\frac{\beta^3  \hbar^2}{6m^2}
 \sum_{k=1}^N \!^{(k \ne j)}
 \left\{
 \frac{u''_{jk} q_{jk,\alpha}^2}{q_{jk}^2}
-
 \frac{u'_{jk} q_{jk,\alpha}^2}{q_{jk}^3}
 +
 \frac{u'_{jk}}{ q_{jk}}
\right\}
 \nonumber \\ & \equiv &
 \sum_{k \in {\cal N}_j} B^{(j)}_{k \alpha} .
\end{eqnarray}
Here ${\cal N}_j$ is a list of the bosons within the potential cut-off
neighborhood of $j$.
The quantity $B^{(j)}_{k \alpha}$ is the contribution of boson $k$
to $B_{j\alpha}$,
and it can be used to calculate the change in the latter with
a change in ${\bf q}_k$,
and \emph{vice versa} for $B^{(k)}_{j \alpha}$.
The quantity $B_{j\alpha}$ will combine with the classical kinetic energy
to give an effective inverse temperature for each boson momentum component,
$\beta_{j\alpha} = \beta - 2 m B_{j\alpha}$.

The imaginary part  is
\begin{eqnarray}
W^{(3)}_{\rm i}
& = &
\frac{-\beta^3}{3!}
\frac{ \hbar^3}{m^2}  {\bf p} \!\cdot\! \nabla \nabla^2 U
 \\ & = &
\frac{-\beta^3\hbar^3}{3!m^2}
\sum_{j,k} {\bf p}_j \cdot \nabla_j \nabla^2 u_{jk}
\nonumber \\ & = &
\frac{-\beta^3\hbar^3}{3!m^2}
\sum_{j,k} {\bf p}_j \cdot \nabla_j [\nabla_j^2+\nabla_k^2] u_{jk}
\nonumber \\ & = &
\frac{-2\beta^3\hbar^3}{3!m^2}
\sum_{j,k} {\bf p}_j \cdot \nabla_j \nabla_j^2  u_{jk}
\nonumber \\ & = &\nonumber
\frac{-2\beta^3\hbar^3}{3!m^2}
\sum_{j,k}\!^{( k \ne j)}
\!\left[ \frac{u'''_{jk}}{q_{jk}}
+ \frac{2u''_{jk}}{q_{jk}^2}
- \frac{2u'_{jk}}{q_{jk}^3}\right] {\bf p}_j \cdot {\bf q}_{jk}.
\end{eqnarray}
This contribution  to the weight is
\begin{eqnarray}
e^{ {\rm i} W^{(3)}_{\rm i}({\bf p},{\bf q}) }
& = &
\prod_{j,k} e^{  ({- {\rm i}\beta^3 \hbar^3}/{3 m^2} )
\left[ \frac{u'''_{jk}}{q_{jk}}
+ \frac{2u''_{jk}}{q_{jk}^2}
- \frac{2u'_{jk}}{q_{jk}^3}\right]
{\bf p}_j \cdot {\bf q}_{jk}
}
\nonumber \\ & \equiv &
\prod_{j}
e^{ -{\rm i} {\bf b}_{j} \cdot {\bf p}_{j}/\hbar } ,
\end{eqnarray}
where the third order `position' is
\begin{equation}
{\bf b}_{j}({\bf q}) \equiv
\frac{\hbar^4\beta^3 }{3 m^2 }
\sum_{k=1}^N \!^{(k \ne j)}
\left[ \frac{u'''_{jk}}{q_{jk}}
+ \frac{2u''_{jk}}{q_{jk}^2}
- \frac{2u'_{jk}}{q_{jk}^3} \right]
{\bf q}_{jk}  .
\end{equation}

Everything goes through as for the pair term
with $\beta \Rightarrow \beta_{j\alpha}$,
 $\Lambda \Rightarrow \Lambda_{j\alpha}$
and
$ {\bf a}_{j} \Rightarrow {\bf a}_{j}+{\bf b}_{j}$.
Thus the momentum integral for the average
of the second and third order terms is
\begin{eqnarray}
I_{W3} & = &
\frac{1}{h^{3N}} \int {\rm d}{\bf p}\;
e^{-\beta p^2/2m}
\nonumber \\ && \mbox{ } \times
\prod_{j,\alpha} \left[
e^{ -{\rm i}{a}_{j\alpha} { p}_{j\alpha}/\hbar }
e^{B_{j\alpha} p_{j\alpha}^2}
e^{ -{\rm i}{b}_{j\alpha} { p}_{j\alpha}/\hbar }
\right]
\nonumber \\ & = &
 \prod_{j,\alpha}
\left[ \Lambda_{j\alpha}^{-1 }
e^{ -\pi c_{j\alpha}^2 /\Lambda_{j\alpha}^2 }
\right] .
\end{eqnarray}
Here and below
${c}_{j\alpha} \equiv {a}_{j\alpha} +{b}_{j\alpha} $,
$\beta_{j\alpha} \equiv \beta - 2m B_{j\alpha}$,
and $\Lambda_{j\alpha} \equiv \sqrt{2\pi\hbar^2 \beta_{j\alpha}/m}$.

In summary,
the position configuration space weight
in the third order diagonal approximation is
\begin{equation}
\wp^{(3)}_{\rm diag}({\bf q})
 =
\frac{ e^{-\beta U({\bf q})} e^{\tilde W_{\rm r}({\bf q})} }{N!Z }
\prod_{j,\alpha}
\frac{e^{ -\pi c_{j\alpha}({\bf q})^2 /\Lambda_{j\alpha}({\bf q})^2 }
}{ \Lambda_{j\alpha}({\bf q})  }.
\end{equation}
Here $\tilde W_{\rm r} = \tilde W_{\rm r}^{(2)}
+ \tilde W_{{\rm r},a}^{(3)}+\tilde W_{{\rm r},b}^{(3)}$.
The partition function is
\begin{eqnarray}
\lefteqn{
Z(N,V,T)
} \\ \nonumber
& = &
\frac{1}{N!}
\int {\rm d}{\bf q}\;
e^{-\beta U({\bf q})} e^{\tilde W_{\rm r}({\bf q})}
\prod_{j,\alpha}
\frac{e^{ -\pi c_{j\alpha}({\bf q})^2 /\Lambda_{j\alpha}({\bf q})^2 }
}{ \Lambda_{j\alpha}({\bf q})  }.
\end{eqnarray}

\subsection{Metropolis Monte Carlo Step}

The classical Metropolis Monte Carlo algorithm for pair potentials
can be very efficient if neighbor tables are used,
in which case it scales linearly with the number of particles $N$.
Some, but not all, of the terms in the present position-dependent
commutation function are the sum of pair terms,
and so they fit naturally into existing schemes.
Careful attention has to be paid to the term that
is the product of gradients of the potential,
and to the momentum-derived terms
in order to compute the Metropolis Monte Carlo step efficiently.

We can write ${\bf c}_j
\equiv {\bf a}_j + {\bf b}_j
= \sum_k^{(k \ne j)} f_{jk} {\bf q}_{jk}$
where
\begin{equation} \label{Eq:fjk}
f_{jk} =
\frac{ \beta^2\hbar^2  }{ 2 m }
\frac{ u'_{jk} }{ q_{jk}}
+
\frac{\beta^3\hbar^4 }{3 m^2 }
\left[ \frac{1}{q_{jk}} u'''_{jk}
+ \frac{2}{q_{jk}^2}u''_{jk}
- \frac{2}{q_{jk}^3}u'_{jk} \right] .
\end{equation}
The momentum-integrated Gaussian exponent is
\begin{equation}
W_p({\bf q}) =
-\pi \sum_{j,\alpha} \frac{c_{j\alpha}^2}{\Lambda_{j\alpha}^{2}} .
\end{equation}
For a trial move ${\bf q}_j \Rightarrow {\bf q}_j'$,
the change in this is
\begin{eqnarray}
\Delta_{j\alpha} W_{p}
& = &
\frac{-\pi c_{j\alpha}'^2}{\Lambda_{j\alpha}'^2}
+
\frac{\pi c_{j\alpha}^2}{\Lambda_{j\alpha}^2}
\nonumber \\ && \mbox{ }
- \pi \sum_{k \in {\cal N}_j}
\left\{
\frac{ c_{k\alpha}'^2 }{ \Lambda_{k\alpha}'^2 }
-
\frac{ c_{k\alpha}^2 }{ \Lambda_{k\alpha}^2 }
 \right\} .
\end{eqnarray}
Here ${\cal N}_j$ is a list of the bosons within the potential cut-off
neighborhood of $j$.
We have
\begin{equation}
c_{k\alpha}'
=
c_{k\alpha} + f_{jk} {q}_{jk,\alpha}' -  f_{jk} {q}_{jk,\alpha},
\end{equation}
and
\begin{equation}
\Lambda_{k\alpha}'^2
=
\frac{2\pi\hbar^2}{m}
\left\{ \beta
- 2m \left[ B_{k\alpha} + B'^{(k)}_{j\alpha} -  B^{(k)}_{j\alpha} \right]
\right\} .
\end{equation}
These require ${\cal O}({\cal N}_j)$ terms to be computed for each change
in  position, which is the same as for the pair potential.

The change in the
pre-exponential factor of the weight,
$\prod_{j,\alpha} \Lambda_{j,\alpha}({\bf q})^{-1}$
also follows from this last result.

Similar techniques can be used for the third order real term
that is quadratic in the potential.
The change in this for ${\bf q}_j \Rightarrow {\bf q}_j'$ is
\begin{eqnarray}
\Delta_j \tilde W^{(3)}_{{\rm r},a}
& = &
\frac{\beta \hbar^2}{3!m }
\left\{ \sum_\alpha [ g_{j\alpha}'^2- g_{j\alpha}^2]
\right.  \\ && \left. \mbox{ }\nonumber
+
\sum_{k \in {\cal N}_j} \sum_\alpha
\left[
\left( g_{k\alpha} +  g^{(k)}_{j\alpha}\!\,' -  g^{(k)}_{j\alpha} \right)^2
- g_{k\alpha}^2 \right]
\right\}.
\end{eqnarray}
Here the $j$-dependent contribution to ${\bf g}_k$ is
\begin{equation}
{\bf g}^{(k)}_j
=  \frac{u_{jk}'}{q_{jk}} {\bf q}_{kj}.
\end{equation}
Again it takes on the order of ${\cal N}_j$ terms to evaluate the change,
which is the same as for the pair potential itself.

\subsection{Average Energy and Heat Capacity}

As above we write ${\bf c}_j
\equiv {\bf a}_j + {\bf b}_j
= \sum_k^{(k \ne j)} f_{jk} {\bf q}_{jk}$,
where  the above expression shows that
$  f_{jk} \equiv  \beta^2  f_{jk}^{(2)}  + \beta^3  f_{jk}^{(3)}$.
Using an over-dot to denote the inverse temperature derivative,
we have
$\dot f_{jk} = 2 \beta  f_{jk}^{(2)} + 3 \beta^2 f_{jk}^{(3)}$
and
$\ddot f_{jk} = 2  f_{jk}^{(2)} + 6 \beta f_{jk}^{(3)}$.
Hence $\dot c_{j\alpha}$ and $\ddot c_{j\alpha}$ follow.
Since $B_{j\alpha} \propto \beta^3$,
$\dot B_{j\alpha} = 3 B_{j\alpha}/\beta$
and $\ddot B_{j\alpha} = 6 B_{j\alpha}/\beta^2$.
And since
$ \beta_{j\alpha}({\bf q})
\equiv   \beta - 2mB_{j\alpha}({\bf q}) $,
then $\dot \beta_{j\alpha} = 1 - 2m \dot B_{j\alpha}$,
and $\ddot \beta_{j\alpha} =  - 2m \ddot B_{j\alpha}$.
The derivatives of
$\Lambda_{j\alpha}({\bf q}) = \sqrt{2\pi\beta_{j\alpha}\hbar^2/m}$
also follow.

The negative of the temperature-dependent exponent
of the position configuration probability is
\begin{equation}
{\rm H}^\S ({\bf q}) \equiv
\beta U({\bf q})
- {\tilde W}_{\rm r}({\bf q})
+ \pi \sum_{j,\alpha}
\frac{ c_{j,\alpha}({\bf q})^2 }{ \Lambda_{j,\alpha}({\bf q})^2 }
+ \sum_{j,\alpha} \ln \Lambda_{j,\alpha}({\bf q}) .
\end{equation}
The thermal wavelength that appears as the argument of the logarithm
has the same unit of length as the position configuration,
but this is not important for the derivatives.
The inverse temperature derivative of this is a sort of statistical energy
or enthalpy,
\begin{eqnarray}
\lefteqn{
{\rm H}({\bf q})
 \equiv
\frac{\partial {\rm H}^\S({\bf q}) }{\partial \beta}
} \nonumber \\
& = &
U - \dot{\tilde W}_{\rm r}
+  \sum_{j,\alpha}
\left\{
\frac{ 2\pi \dot c_{j,\alpha}c_{j,\alpha} }{ \Lambda_{j,\alpha}^2 }
- \frac{\pi c_{j,\alpha}^2}{\beta_{j\alpha} \Lambda_{j,\alpha}^2 }
\dot \beta_{j\alpha}
\right. \nonumber \\ && \left. \mbox{ }
+ \frac{\dot \beta_{j\alpha}}{2 \beta_{j,\alpha}}
\right\}.
\end{eqnarray}

The  monomer thermodynamic  energy is
\begin{equation}
\overline E^{(1)}
=
\frac{-\partial \ln Z}{\partial \beta}
=
\left\langle {\rm H} \right\rangle .
\end{equation}

The heat capacity is $C_V = \partial \overline E^{(1)} /\partial T
= - k_{\rm B}\beta^2 \partial \overline E^{(1)} /\partial \beta$.
We have
\begin{equation}
\frac{\partial \overline E^{(1)}}{\partial \beta}
=
\left\langle {\rm H} \right\rangle^2
-
\left\langle  {\rm H}^2 - \dot {\rm H} \right\rangle
=
\left\langle \dot {\rm H}
- \left[ {\rm H}-\left\langle {\rm H} \right\rangle \right]^2  \right\rangle ,
\end{equation}
with
\begin{eqnarray}
\dot {\rm H}
&  = &
- \ddot{\tilde W}_{\rm r}
+  \sum_{j,\alpha}
\left\{
\frac{ 2 \pi\ddot c_{j,\alpha}c_{j,\alpha} +  2\pi\dot c_{j,\alpha}^2
}{ \Lambda_{j,\alpha}^2 }
-
\frac{ 2 \pi\dot c_{j,\alpha}c_{j,\alpha}
}{ \beta_{j,\alpha} \Lambda_{j,\alpha}^2 }\dot \beta_{j\alpha}
\right. \nonumber \\ && \left. \mbox{ }
-
\frac{2\pi\dot c_{j,\alpha}c_{j,\alpha}
}{\beta_{j\alpha} \Lambda_{j,\alpha}^2 }\dot \beta_{j\alpha}
+
\frac{2\pi c_{j,\alpha}^2}{\beta_{j\alpha}^2 \Lambda_{j,\alpha}^2 }
\dot \beta_{j\alpha}^2
- \frac{\pi c_{j,\alpha}^2}{\beta_{j\alpha} \Lambda_{j,\alpha}^2 }
\ddot \beta_{j\alpha}
\right. \nonumber \\ && \left. \mbox{ }
+ \frac{\ddot \beta_{j\alpha}}{2 \beta_{j,\alpha}}
- \frac{\dot\beta_{j,\alpha}^2}{2 \beta_{j,\alpha}^2}
\right\} .
\end{eqnarray}

\subsection{Kinetic Energy}

The kinetic energy
as a function of the position configuration
in the third order, diagonal approximation is
\begin{equation} \label{Eq:KE}
{\cal K}({\bf q})
=
\frac{1}{2} \sum_{k=1}^N \sum_{\gamma=x,y,z}  \beta_{k\gamma}({\bf q})^{-1} .
\end{equation}
The average of this is expected to be less
than the classical average kinetic energy,
$\overline {\cal K}^{\rm cl} = 3N /2\beta$.

\subsection{Symmetrization Function}

It is a relatively minor matter to include the symmetrization function
in the present third order diagonal approximation.
It simply requires the replacement $\Lambda \Rightarrow \Lambda_{j\alpha}$.
From earlier work  (Attard 2018b, 2021),
the symmetrization $l$-loop  grand potential is given by
\begin{equation}
-\beta \Omega^{(l)}
=
\left\langle \eta^{(l)}({\bf \Gamma}) \right\rangle ,
\end{equation}
where the $l$-loop symmetrization function is
\begin{equation}
\eta^{(l)}({\bf \Gamma})
=
\sum_{j_1,\ldots,j_l}^N\!\!\!'\; \prod_{k=1}^{l}
e^{ - {\bf p}_{j_k} \cdot {\bf q}_{j_k,j_{k+1}} /{\rm i}\hbar}
 , \quad j_{l+1} \equiv j_1 .
\end{equation}
The sum is over the unique directed cyclic permutations
of all subsets of $l$-bosons.
Using the third order  diagonal approximation
the momentum quadrature reduces this to
\begin{equation}
\eta^{(l)}({\bf q})
=
\sum_{j_1,\ldots,j_l}^N\!\!\!'\; \prod_{k=1}^{l} \prod_{\alpha=x,y,z}
e^{-\pi ({q}_{j_k j_{k+1},\alpha}-c_{j_k\alpha})^2/\Lambda_{j_k\alpha}^2}
. 
\end{equation}

%
\section{Computational Results}
\setcounter{equation}{0} \setcounter{subsubsection}{0}
\renewcommand{\theequation}{\arabic{section}.\arabic{equation}}
%

\subsection{Model and Simulation Details}

Metropolis Monte Carlo simulations were performed in configuration
position space using the third order diagonal approximation
for the Wigner-Kirkwood commutation function.
The Lennard-Jones pair potential was used,
$u(r) = 4 \varepsilon[(\sigma/r)^{12} - (\sigma/r)^{6} ]$,
with helium parameters,
$ \varepsilon_{\rm He} = 10.22 k_{\rm B}$\,J and
$\sigma_{\rm He} =0.2556$\,nm (van Sciver 2012).
The potential was cut off at $R_{\rm cut} = 3.5\sigma$.
Periodic boundary conditions were used
together with the nearest neighbor convention.
A small cell spatial neighbor table was used.
The position steps for a trial configuration
were chosen to give acceptance rates of 30--60\%.
Usually $N=1,000$ atoms were used,
and these formed a homogeneous fluid.


\subsection{Results}

\subsubsection{Numerical Momentum Quadrature}

\begin{table}[tb]
\caption{
Quantum Monte Carlo in classical phase space
with numerical momentum quadrature (Attard 2025h).
Results for Lennard-Jones $^4$He
using $N=1,000$, full momentum quadrature, $n_w^{\rm max}=3$,
and $|W_{\rm i}| < \pi/2$, unless otherwise indicated.
The statistical error in the final digit shown in parentheses
is at the 95\% confidence level.
\label{Tab:QMC1} }
\begin{center}
\begin{tabular}{c c c c c c c}
\hline\noalign{\smallskip}
$k_\mathrm{B} T /\varepsilon $ & $\rho\sigma^3$ &
$ \beta \overline {\cal K}/N $ &
$ \beta \overline E/N$  &
$C_V/Nk_{\rm B}$      \\
\hline 
0.50 & 0.26 & 0.8538(2) &  $-10.988(2)$ & 28.8(3)  \\
0.50$^a$ & 0.26 & 0.8544(2) &  $-10.983(4)$ & 29.0(5)  \\
0.50$^b$ & 0.26 & 0.850(9) &  $-11.00(6)$ & ---  \\
0.50$^c$ & 0.26 & 0.6508(2) &  $-11.947(4)$ & 27.2(3)  \\
0.50$^d$ & 0.26 & 0.455(1) &  $28.7(5)$ & $-69(11)$  \\
0.49 & 0.26 & 0.8429(4) &  $-11.816(4)$ & 30.5(3)  \\
0.49$^d$ & 0.26 & 0.441(1) &  $30.8(6)$ & $-81(2)$  \\
0.45 & 0.26 & 0.7935(3) &  $-15.917(6)$ & 38.9(4)  \\
0.40 & 0.26 & 0.7225(7) &  $-23.67(1)$ & 55.1(9)  \\
0.35 & 0.26 & 0.633(2) &  $-36.9(1)$ & 120(25)  \\

\hline
\end{tabular}
\end{center}
\flushleft
$^a$$N=2,000$.
$^b$Umbrella sampling, $e^{-W_{\rm i}^2/2}$.
$^c$Diagonal approximation.
$^d$$n_w^{\rm max}=4$.
\end{table}

Table~\ref{Tab:QMC1} provides benchmark results obtained with the previous
computer simulation algorithm in which the momentum integrals
were evaluated by Metropolis Monte Carlo quadrature.
These results supersede those in versions 1 and 2 of Attard (2025h).

The result are for a homogeneous fluid of density $\rho \sigma^3 = 0.26$.
This is the liquid saturation density
of $^4$He at $k_{\rm B}T/\varepsilon =0.5$,
as estimated from the measured data (Donnelly and Barenghi, 1998).
The second entry shows that the algorithm is relatively insensitive
to the size of the system.

The umbrella sampling method removes the constraint $|W_{\rm i}| < \pi/2$
and replaces the weight $\cos W_{\rm i}$
with $e^{-W_{\rm i}^2/2}$ in the Metropolis Monte Carlo step,
and corrects for this in the averaging step
with the factor  $e^{W_{\rm i}^2/2}\cos W_{\rm i}$.
This comes at the cost of greater statistical error,
most markedly for the heat capacity where it was about 4 orders of magnitude
larger than for direct sampling for the same length simulation.
Nevertheless
the agreement of the umbrella sampling results with the direct results
for the kinetic energy and for the thermodynamic energy
suggests that the constraint  $|W_{\rm i}| < \pi/2$
has no deleterious consequences.

The results at $k_\mathrm{B} T /\varepsilon = 0.49$
give the numerical derivative
$\Delta \overline E/\Delta T = 29.6(2)$,
which is in good agreement with the heat capacity simulated directly,
which was obtained using the formula in version 1 of Attard (2025h).
This is a sensitive test for computer bugs.
It also confirms that the results are insensitive
to the constraint $|W_{\rm i}| < \pi/2$.

The fourth entry in Table~\ref{Tab:QMC1}
shows that the diagonal approximation is relatively accurate.
For these results the momenta still appeared explicitly
and the momentum integrals were still evaluated
using the Metropolis Monte Carlo method.
The only difference was that Eq.~(\ref{Eq:pdiag}) was used
instead of the full expression.
The approximation underestimates the kinetic energy
by about 25\%,
overestimates the magnitude of the energy by about 10\%,
and underestimates the heat capacity by about 5\%.
It can be concluded that this approximation is relatively reliable.

The fifth and seventh entries in Table~\ref{Tab:QMC1}
use the fourth order expansion, $n_w^{\rm max}=4$.
In these simulations
there were indications of slow equilibration and of solidification,
and the diffusion over their course was extremely limited,
with the average change in separation of a pair of atoms being $0.1\sigma$.
The thermodynamic phase of Lennard-Jones $^4$He is sensitive
to the order of the expansion for the commutation function,
and so the quantitative results at this density and temperature
should be treated with caution.
It is problematic to compare these fourth order results
that yield a solid-like phase
with the third order results that give a fluid phase.
For what it is worth,
it can be seen that at $k_{\rm B}T /\varepsilon = 0.50$,
compared to the third order expansion,
the kinetic energy is substantially smaller
and the statistical enthalpy is substantially more positive.
The heat capacity given by the fourth order expansion
is large in magnitude and negative in sign,
which indicates that the state is thermodynamically unstable.
The numerical difference in thermodynamic energy for these two states was
$\Delta \overline E/\Delta T = -75(38)$,
which is consistent with the heat capacities obtained
by direct simulation.
Runs at  $k_{\rm B}T /\varepsilon = 0.50$ and $\rho\sigma^3=0.20$
appear to have equilibrated as solids
with $\beta \overline {\cal K}/N=0.617(2)$,
$\beta \overline E/N=-4.6(1)$,
and $C_V/Nk_{\rm B} = 36.3(14)$.

One can conclude that
at these temperatures and densities
Lennard-Jones $^4$He is close to coexisting solid, liquid, and vapor phases.
One therefore expects that the structure and the dynamics of the system
would be sensitive to the details of the model intermolecular potential
and to the order of the expansion
for the Wigner-Kirkwood commutation function.
The proximity of model $^4$He to the phase boundaries
makes it challenging to obtain reliable quantitative results
that can be compared to real $^4$He.

\subsubsection{Analytic Diagonal Approximation}

\begin{table}[tb]
\caption{
Simulation results 
in position configuration space
using the third order diagonal analytic approximation for the momentum.
\label{Tab:QMC2} }
\begin{center}
\begin{tabular}{c c c c c c c c c}
\hline\noalign{\smallskip}
$k_\mathrm{B} T /\varepsilon $ & $\rho\sigma^3$ & $\Lambda/\sigma$ &
$\overline \Lambda_{j\alpha}/\sigma $  &
$ \beta \overline {\cal K}/N $ &
$ \beta \overline E/N$  &
$C_V/Nk_{\rm B}$      \\
\hline 
1.0  & 0.26 & 1.0679 & 1.244(1) & 1.105(2) & 1.0745(8) & 4.98(5)\\
0.9  & 0.26 & 1.1257 & 1.350(2) & 1.043(3) & 0.5804(5) & 6.11(9)  \\
0.8  & 0.26 & 1.1940 & 1.486(1) & 0.968(1) & -0.226(1) & 8.0(1) \\
0.7  & 0.26 & 1.2764 & 1.667(2) & 0.880(3) & -1.613(2) & 11.10(4) \\
0.6  & 0.26 & 1.3787 & 1.921(2) & 0.772(2) & -4.160(1) & 16.8(1) \\
0.5  & 0.26 & 1.5103 & 2.300(2) & 0.6467(9) & -9.291(2)  & 27.3(2)\\
0.49 & 0.26 & 1.5256 & 2.349(3) & 0.629(2) & -10.054(2) & 29.2(1)\\
0.45 & 0.26 & 1.5920 & 2.570(1) & 0.575(1) & -13.850(3) & 36.6(1) \\
0.4  & 0.26 & 1.6886 & 2.929(3) & 0.498(4) & -20.987(3) & 50.8(1) \\
0.35 & 0.26 & 1.8051 & 3.42(2)  & 0.422(8) & -32.76(3)  & 78.0(9) \\
\hline
\end{tabular}
\end{center}
\flushleft
\end{table}

Table~\ref{Tab:QMC2} shows the results of the present
quantum Monte Carlo simulations
in third order diagonal approximation
in position configuration space.
The fixed density $\rho \sigma^3 = 0.26$,
which is the liquid saturation density at $k_\mathrm{B} T /\varepsilon = 0.5$,
underestimates the saturation density at lower temperatures,
being about 25\% less  than the measured saturation value
at the lowest temperature simulated,
$\rho^{\rm sat} \sigma^3 = 0.34$ at $k_{\rm B}T/\varepsilon =0.35$.
An attempt was made to use the saturation density
measured at each temperature,
but it was found that the system solidified
for $k_{\rm B}T/\varepsilon \le 0.45$.
For all temperatures at $\rho \sigma^3 = 0.26$,
the system appeared to be a stable fluid,
except for the lowest temperature, $k_{\rm B}T/\varepsilon =0.35$,
where the simulated system was a metastable fluid
that eventually transformed to a solid for longer simulations.

It can be seen in Table~\ref{Tab:QMC2}
that the effective thermal wavelength
$\overline \Lambda_{j\alpha}/\sigma $
averaged over all particles and components
is larger than the classical value,
increasingly so as the temperature decreases.
It is almost twice the size of the classical thermal wavelength
at the lowest temperature studied.
This is a manifestation of the fact that the effective temperature
in the quantum system is less than the applied temperature.
This is an unusual and unexpected consequence
of the Heisenberg uncertainty relation
as manifest in the Wigner-Kirkwood commutation function.
Since the $\lambda$-transition occurs when $\rho \Lambda^3 ={\cal O}(1)$,
that the effective thermal wavelength is larger than the classical value
means that the transition will occur at higher temperatures
than would have been predicted on the basis of classical considerations.

A closely related quantity is the kinetic energy per particle,
which in  Table~\ref{Tab:QMC2} can be seen to be much less
than the classical value, $\beta {\cal K}^{\rm cl}/N = 3/2$.
Again this can be interpreted as the effective temperature
of the quantum system being much less than the reservoir temperature.
Alternatively,
one can say that the bosons occupy lower momentum states
than would be predicted by classical considerations.

The kinetic energy per particle listed in Table~\ref{Tab:QMC2}
is actually $3 \Lambda^2/2\overline \Lambda_{j\alpha}^2$.
This is approximately the same as the exact result, Eq.~(\ref{Eq:KE}).
At the temperature $k_{\rm B}T/\varepsilon =0.5$,
the result in the table is $\beta {\cal K}/N = 0.6467(9)$,
whereas the result of Eq.~(\ref{Eq:KE}) is 0.66122(5).
This can be compared with the numerical third order diagonal approximation
in  Table~\ref{Tab:QMC1}, namely 0.6508(2),
and with the full numerical benchmark, 0.8538(2).
There is a statistically significant difference of 1--2\%
between the analytic (Table~\ref{Tab:QMC1})
and the numerical (Table~\ref{Tab:QMC2})
third order diagonal approximations, the origin of which is unclear.
Possibly it reflects a limitation of the constraint
$|W_{\rm i}|<\pi/2$ that is applied in the numerical quadrature,
but not in the analysis.

Table~\ref{Tab:QMC2} also shows
the average thermodynamic energy and the specific heat capacity
from the simulations.
The heat capacity is positive, as it must be,
and it increases with decreasing temperature.
This is the opposite trend to that which is measured in the laboratory,
where the heat capacity decreases with decreasing temperature
prior to diverging at the the $\lambda$-transition.
(The present results are on a constant density curve,
whereas the experimental results are on the saturation curve.)
The lowest temperature in the table is about 1.4\,K
above the $\lambda$-transition at $T_\lambda = 2.18$\,K.
Since symmetrization effects are not here included,
the present increase in heat capacity with decreasing temperature
is unlikely to have anything to do with the measured divergence
at the $\lambda$-transition.
Instead it likely reflects the approach to the liquid-solid transition
in the simulated system.

The minimum specific heat capacity on the saturation curve
prior to the $\lambda$-transition is 9.04\,J/mol\,K at 2.55\,K
(Donnelly and Barenghi 1998).
In the present dimensionless units this is  $C_{\rm s}/Nk_{\rm B} = 1.09$
at $k_{\rm B}T/\varepsilon = 0.25$.
This is much lower than the simulation results in Table~\ref{Tab:QMC2}.
It is not expected that the difference between
the heat capacity at constant saturation
and the heat capacity at constant volume
could account for this discrepancy.
It is more likely that the larger than expected heat capacity
is due to the proximity
to the liquid-solid transition in the present model.

The present results for the third order analytic diagonal approximation
for $k_{\rm B}T/\varepsilon=0.50$ and  $\rho\sigma^3=0.26$
are
$\beta E/N = -9.292(2)$ and $-\beta^2 \dot E/N = 27.4(2)$.
For $k_{\rm B}T/\varepsilon=0.49$ and  $\rho\sigma^3=0.26$,
$\beta E/N = -10.054(2)$ and $-\beta^2 \dot E/N = 29.2(1)$
(Table~\ref{Tab:QMC2}).
From these energies,
the numerical derivative is
$\Delta E /Nk_{\rm B}\Delta T = 28.0(2)$.
That this agrees with the directly simulated heat capacity
is a very sensitive test of the mathematical expressions
for the energy and the heat capacity,
and of their computer implementation.
It also tests the Metropolis Monte Carlo algorithm
and its implementation.
The author can attest to the sensitivity of the procedure
as it took many rounds of debugging
to get quantitative agreement between the two.

The agreement of the present results for the energy and the heat capacity
with the earlier more exact simulations
tends to confirm the utility of the diagonal approximation.
This conclusion must be balanced against the disagreement
in the kinetic energy pinpointed above.

\begin{figure}[t]
\centerline{ \resizebox{8cm}{!}{ \includegraphics*{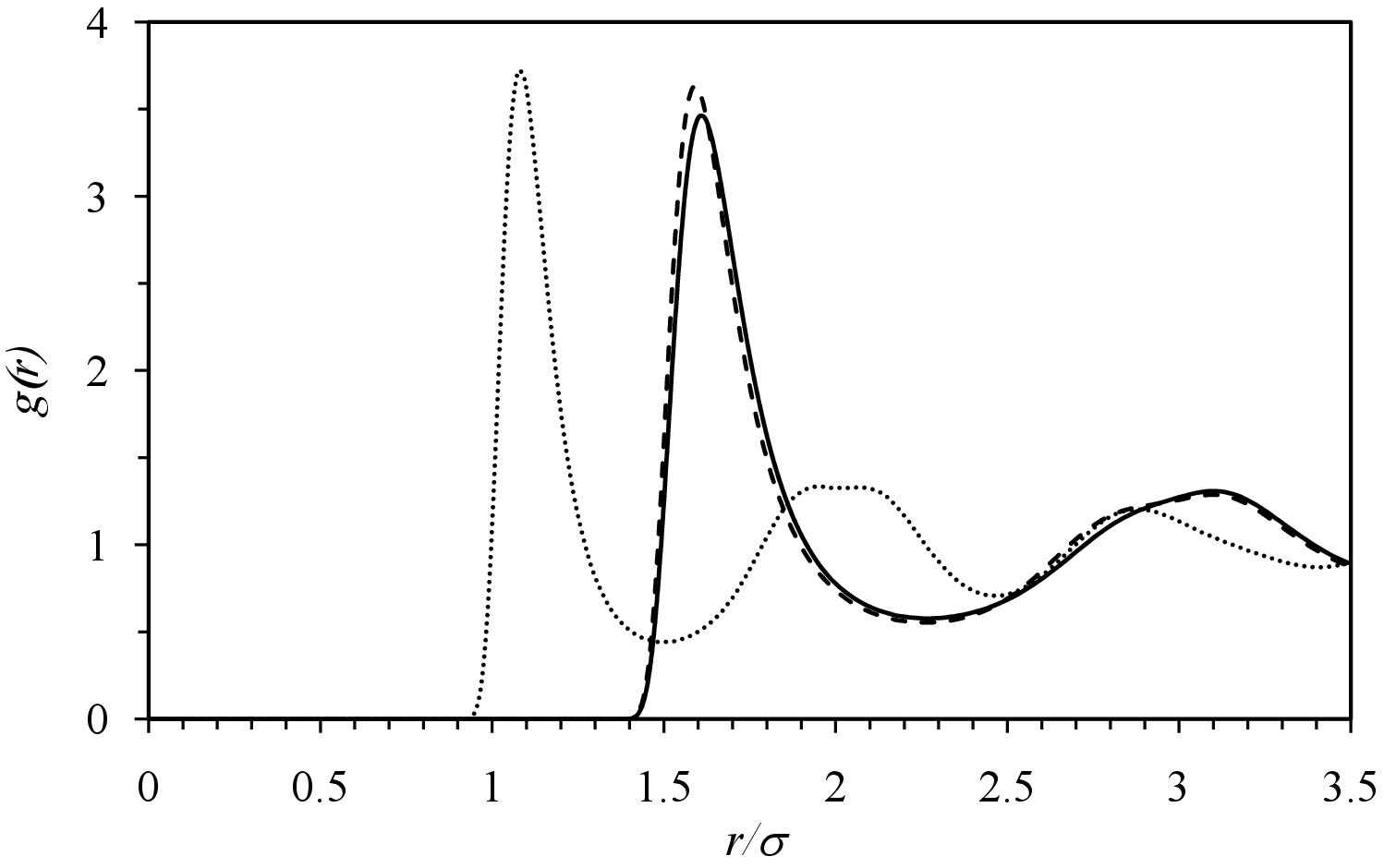} } }
\caption{\label{Fig:g(r)}
Radial distribution function for Lennard-Jones $^4$He
at $k_{\rm B}T/\varepsilon =0.5$ and $\rho\sigma^3=0.26$.
The solid curve is the present third order diagonal approximation,
the dashed curve is the full third order with numerical momentum quadrature,
and the dotted curve is the classical result at $\rho\sigma^3=0.9331$.
}
\end{figure}

Figure~\ref{Fig:g(r)}
compares the radial distribution function
obtained with the present third order diagonal approximation
with that obtained with the numerical full momentum quadrature.
It can be seen that the two are in relatively close agreement.
This suggests that for structural properties at least
the diagonal approximation is quite good.
This possibly explains
the good agreement for the energy and the heat capacity
discussed above.
Also shown in the figure is the classical result
at the higher density $\rho\sigma^3=0.9331$,
which is the classical Lennard-Jones $^4$He saturation density
at $k_{\rm B}T/\varepsilon =0.5$.
The most noticeable change is the increases
in the size of the core or exclusion region
in the quantum system.
This is evidently a direct result of the Wigner-Kirkwood commutation function,
and it can be interpreted as a manifestation
of the Heisenberg uncertainty relation.

A better measure
of the location of the $\lambda$-transition
than $\rho \Lambda^3 = {\cal O}(1)$
is arguably $\overline \Lambda_{j\alpha} /r_{\rm peak} \agt 1$,
where $r_{\rm peak}$ gives the peak of the radial distribution function.
(Or
$\rho g_{\rm peak} 2\pi r_{\rm peak}^2 \Delta_{\rm peak}
e^{-\pi r_{\rm peak}^2 / \overline \Lambda_{j\alpha}^2}  \agt 1$.)
For the classical Lennard-Jones fluid,
$r_{\rm peak}^{\rm cl}  \alt 2^{1/6}\sigma = 1.1\sigma$,
and $\Lambda/r_{\rm peak}^{\rm cl} = 1.5 /1.1 = 1.4$
at $k_{\rm B}T/\varepsilon =0.5$.
In the quantum case in Fig.~\ref{Fig:g(r)},
$r_{\rm peak} = 1.58 \sigma $, $g_{\rm peak}=3.62$,
and $\overline \Lambda_{j\alpha} = 2.3\sigma$,
giving  $\overline \Lambda_{j\alpha}/r_{\rm peak} = 1.4$.
This is about the same as the classical estimate.
(For $n_w^{\rm max}=4$, $r_{\rm peak} = 1.69 \sigma $
and $g_{\rm peak}=6.90$.)
Of course this is a rather crude estimate of the transition,
and the two quantities have different rates of change with temperature.
Nevertheless it suggests that the commutation function
contributes two competing effects in locating the $\lambda$-transition:
the lower effective temperature and larger effective thermal wavelength
makes it more favorable,
whereas the increased separation of closest approach between pairs of bosons
makes it less favorable.

It should be mentioned that a hard-core potential was imposed
with diameter inside the region in which the radial distribution function
was anyway zero.
It had no effect on the results.
For example, the results at $k_{\rm B}T/\varepsilon = 0.50$,
$\rho \sigma^3=0.26$
with the hard-core diameter set at $1.40\sigma$ and at $1.28\sigma$
were statistically indistinguishable.
The results in Table~\ref{Tab:QMC1} were obtained with
the hard-core diameter set at $1.18\sigma$.

%
\section{Conclusion}
\setcounter{equation}{0} \setcounter{subsubsection}{0}
\renewcommand{\theequation}{\arabic{section}.\arabic{equation}}
%

The main point of this paper is to explore
an approximation that reduces the complex phase space weight
for a quantum system to a real weight in position configuration space.
This is called the third order diagonal approximation,
referring to the termination of the series for the Wigner-Kirkwood
commutation function
and to the projection of the quadratic momentum terms therein.
The advantage is that the momentum integrals
can be performed analytically rather than numerically
as in the full third order approach.

For Lennard-Jones $^4$He liquid below about 10\,K,
the subsequent Metropolis Monte Carlo simulation algorithm
is found to be tractable.
The algorithm and model seem to more readily solidify than real $^4$He,
and so results were obtained at a constant density
that was about 25\% less than the saturated liquid density at the
lowest temperature studied, $T = 3.5$\,K.
The kinetic energy at $k_{\rm B}T/\varepsilon = 0.5$ ($T=5.1$\,K)
is about 25\% lower
than for the full third order numerical momentum quadrature approach.
The energy, heat capacity, and radial distribution function
have smaller error at this temperature and density.

That the present approach largely agrees with the earlier approach,
and also the umbrella sampling result in Table~\ref{Tab:QMC1},
tends to confirm the treatment
of the complex  phase space weight for a quantum system
that was the basis of the original algorithm (Attard 2025h).

The present results were run for the same number of Monte Carlo steps
as the earlier results,
but took about 50\% longer in computer time.
Since the statistical error is about the same,
we conclude that the present algorithm is slightly less efficient
than the more exact numerical momentum quadrature algorithm.

Despite this,
the analytic momentum quadrature of the third order diagonal approximation
does offer a certain physical insight into the effects
of the Wigner-Kirkwood commutation function,
which embodies the Heisenberg uncertainty relation.
It shows the link between momentum  and position configuration
that is missing in classical statistical mechanics.
It shows quantitatively how the Heisenberg uncertainty relation increases
the distance of closest approach of two bosons,
and how the position configuration lowers
the kinetic energy and the momentum state
compared to the classical prediction.


One significant challenge with the classical phase space algorithm
for the quantum Monte Carlo simulation of $^4$He
is that the model and level of approximation
too readily produce solid phases at low temperatures.
This precludes the exploration of the $\lambda$-transition
in the saturated liquid.
Of course the proximity to a solid phase is not unrealistic
as real  $^4$He does become solid
at low temperatures and elevated pressures,
but it does frustrate the main goal of the simulations.

The problem appears to arise from the approximate nature
of the short ranged, $r^{-12}$, repulsion
in the Lennard-Jones pair potential,
which is magnified with each successive gradient
in the terms in the Wigner-Kirkwood commutation function.
These shift the first peak in the radial distribution function
to larger separations and create an exclusion hole about each atom
that greatly exceeds that given classically by the Lennard-Jones potential.
In the case of the fourth order approximation
to the Wigner-Kirkwood commutation function,
even at half the measured saturation liquid density
this causes solidification.
The Lennard-Jones pair potential
may well be optimized for the gas phase and for classical treatments,
but in the quantum liquid regime a more reliable pair potential is called for.
In this regard the pair potentials explored by
Aziz {\em et al.}\ (1992)
and used by  Ceperley  (1995),
which were obtained from quantum calculations and gas phase measurements,
may be more reliable.

\section*{References}


\begin{list}{}{\itemindent=-0.5cm \parsep=.5mm \itemsep=.5mm}

\item 
Attard P 2016b
Quantum statistical mechanics as an exact classical expansion with
results for Lennard-Jones helium
arXiv:1609.08178v3

\item 
Attard P 2018b
Quantum statistical mechanics in classical phase space. Expressions for
the multi-particle density, the average energy, and the virial pressure
arXiv:1811.00730

\item 
Attard P  2021
\emph{Quantum Statistical Mechanics in Classical Phase Space}
(Bristol: IOP Publishing)

\item 
Attard P 2025a
\emph{Understanding Bose-Einstein Condensation,
Superfluidity, and High Temperature Superconductivity}
(London: CRC Press)

\item 
Attard P 2025b
The molecular nature of superfluidity: Viscosity of helium from quantum
stochastic molecular dynamics simulations over real trajectories
arXiv:2409.19036v5

\item
Attard P 2025h
Quantum Monte Carlo in classical phase space with the Wigner-Kirkwood
commutation function. Results for the saturation liquid density of $^4$He
arXiv:2512.09948

\item
Aziz R A,  Slarnan M J, Koide A,  Allnatt A R, and  Meath W J 1992
Exchange-Coulomb potential energy curves for He-He,
and related physical properties
\emph{Mol.\ Phys.}\ {\bf 77} 321

\item 
Ceperley  D M  1995
Path integrals in the theory of condensed helium
\emph{Rev.\ Mod.\ Phys.}\ {\bf 67} 279

\item 
Donnelly R J and  Barenghi C F 1998
The observed properties of liquid helium at the saturated vapor pressure
\emph{J.\ Phys.\ Chem.\ Ref.\ Data} {\bf 27} 1217

\item 
Kirkwood J G 1933
Quantum statistics of almost classical particles
\emph{Phys.\ Rev.}\ {\bf 44}, 31

\item 
van Sciver  S W 2012
\emph{Helium Cryogenics}
(New York: Springer 2nd edition)

\item 
Wigner E 1932
On the quantum correction for thermodynamic equilibrium
\emph{Phys.\ Rev.}\ {\bf 40}, 749

\end{list}

\end{document}